# Technology ready use of single layer graphene as a transparent electrode for hybrid photovoltaic devices

Zhibing Wang[1,2], Conor P. Puls[2], Neal E. Staley[2], Yu Zhang[3,4], Aaron Todd[3], Jian Xu[3]
Casey A. Howsare[5], Matthew J. Hollander[5], and Joshua A. Robinson[5], and Ying Liu[2,*]

[1] *Advanced Photonics Center, Southeast University, Nanjing 210096, PR China*
[2] *Department of Physics and Material Research Institute, Pennsylvania State University, University Park, PA 16802, U.S.A.*
[3] *Department of Engineering Science and Mechanics, Pennsylvania State University, University Park, PA 16802, U.S.A.*
[4] *State Key Laboratory on Integrated Optoelectronics, and College of Electronic Science and Engineering, Jilin University, Changchun 130012, PR China*
[5] *Electro-Optics Center, Pennsylvania State University, University Park, PA 16229, U.S.A.*

**Abstract**: Graphene has been used recently as a replacement for indium tin oxide (ITO) for the transparent electrode of an organic photovoltaic device. Due to its limited supply, ITO is considered as a limiting factor for the commercialization of organic solar cells. We explored the use of large-area graphene grown on copper by chemical vapor deposition (CVD) and then transferred to a glass substrate as an alternative transparent electrode. The transferred film was shown by scanning Raman spectroscopy measurements to consist of >90% single layer graphene. Optical spectroscopy measurements showed that the layer-transferred graphene has an optical absorbance of 1.23% at a wavelength of 532 nm. We fabricated organic hybrid solar cells utilizing this material as an electrode and compared their performance with ITO devices fabricated using the same procedure. We demonstrated power conversion efficiency up to 3.98%, higher than that of the ITO device (3.86%), showing that layer-transferred graphene promises to be a high quality, low-cost, flexible material for transparent electrodes in solar cell technology.

## 1. Introduction

Transparent electrodes are an important part of hybrid photovoltaic devices that have attracted much interest in recent years because they can be fabricated on flexible, light-weight substrates using low-cost methods[1,2]. Currently indium tin oxide (ITO) and fluorine tin oxide (FTO) are commonly used as the transparent electrode in optoelectronic devices[3,4], but both ITO and FTO suffer significant disadvantages – they are relatively expensive materials[5], limited in transparency in the near-infrared region, and mechanically brittle. In particular, their mechanical properties make them unsuitable for devices on flexible substrates. New electrode materials with good transparency, conductivity, and stability are highly desirable for hybrid photovoltaic applications.

Graphene possesses very high charge carrier mobility and mechanical strength, and furthermore is extremely flexible with little optical absorption. These advantages, together with the possibility of low-cost production of large-area graphene, have made graphene attractive for future optoelectronic applications. Recently, the chemical vapor deposition (CVD)[6-9] of graphene on metal substrates such as Ni and Cu was demonstrated. The resulting large-area, continuous, few-layered graphene has been transferred to desirable substrates suitable for application in photovoltaic devices[11]. In the context of hybrid photovoltaic device applications, layer transferred graphene has an additional advantage over ITO as it can achieve work function matching better than that of ITO if treated by UV or modified by pyrene buanoic acid succidymidyl ester[10], desirable for achieving high efficiency.

*Electronic mail: liu@phys.psu.edu

Both chemically exfoliated graphene oxides and layer-transferred graphene have been used previously in hybrid photovoltaic devices. However, their power conversion efficiency (PCE) was shown to be only 0.26% for devices using exfoliated graphene oxides[10] and up to 1.71% for those using layer-transferred graphene[11]. An improved PCE of 0.4% was obtained using CuPc and $C_{60}$ as the photoactive materials[11]. However, the PCE values of these graphene devices were still lower than their ITO counterparts. In this Letter, we report an improved power conversion efficiency of hybrid photovoltaic devices incorporating layer-transferred graphene which is higher than devices based on ITO that were prepared with an identical procedure.

## 2. Materials and Methods

*2.1 Device Design*
We fabricate our hybrid photovolatic devices (shown schematically in Fig. 1a) using either graphene or ITO as the transparent electrode in a design essentially identical to that used previously[10]. Energy levels for materials included in our hybrid devices are depicted in Fig. 1b. The work function of occupied molecular orbitals of poly(3-hexylthiophene) (P3HT), which provides a hole-injection barrier between graphene or ITO, is 4.9eV. Phenyl-C61-butyric acid methyl ester (PCBM) functions as an electron acceptor in the bulk heterojunction solar cells. Poly(3,4–ethylenedioxythiophene): poly(styrenesulfonate) (PEDOT:PSS) is included as a buffer layer only to facilitate the hole injection/extraction. It was demonstrated previously that the neutral PEDOT:PSS middle anode works as an Ohmic hole contact for P3HT. In such a hybrid photovolatic device, electrons in P3HT are collected at the top Al layer while holes are collected at the bottom graphene or ITO electrode. Photon-induced excitons are separated in the active region, with holes drained from PEDOT to the transparent electrode (graphene or ITO). Given that the work function of graphene, 4.7eV, is slightly higher than that of ITO (4.6eV), we expect that the PCE for graphene devices will be higher than those of ITO devices because of the slightly improved work function matching.

*2.2 Graphene growth and transfer*
In this paper, we present the development of CVD synthesis and a roll-to-roll transfer process of the graphene used in the device described above. The CVD synthesis method, using Cu as substrate for growth, is a promising technique for the synthesis of large-area graphene films. The copper substrate was positioned at the center of a quartz tube and heated to 1000°C at a rate of 40°C/min, and then maintained for 20min under a $H_2$/Ar (1000 and 400 sccm, respectively) atmosphere. $CH_4$/$H_2$/Ar gas mixture (CH4:Ar:H2 = 250:1000:4000 sccm) was then introduced into the quartz tube for 10 min, followed by rapid cooling of the sample to room temperature at a rate of 10°C/s with flowing Ar.

The graphene film grown on Cu foil was layer-transferred to a glass substrate. We started by coating the surface of the graphene film with a thin layer of photoresist to minimize contamination and wrinkling. A solution of Transene Copper Etchant CE-100 and water (1:3) was used to etch away the Cu foil. We kept the solution at 80⁰C for about 2 hours to make sure that all Cu was etched away. The graphene film was then transferred to a glass substrate and kept at room temperature for two days to remove water trapped below the graphene. Finally we used acetone and isopropyl alcohol to clean the top surface of the transferred graphene.

*2.3 Graphene Characterization*

In Fig. 2a and b, we show optical images of graphene films transferred onto $SiO_2$/Si and glass substrates. Although the surface of the transferred graphene is relatively uniform, the difference in thermal expansion coefficient for graphene and Cu leads to the formation of wrinkles during the cool-down of the CVD growth. Visible in the images are spots of greater contrast. Raman spectroscopy measurements with a 488 nm activation wavelength carried out on these spots (see below) revealed the same G- and 2D-band peaks for single-layer graphene prepared by mechanical exfoliation but with intensities higher than that from the surrounding graphene area, suggesting possible additional scattering from contamination below the graphene. We also performed atomic force microscopy (AFM) studies on the layer-transferred graphene on $SiO_2$/Si and quartz substrates, as shown in Fig. 2c and d, resolving the same wrinkles and mounds.

In Fig. 2e, we show Raman spectra of a graphene film grown on Cu and of the same film transferred to a quartz substrate. The observed G- and 2D-band peaks prove that the graphene being probed is monolayer and a small D-band peak indicates low disorder. Two-dimensional Raman topographical scans revealed that the film was >90% monolayer. After transfer onto a quartz substrate, the intensity of the D-band peak in the spectrum was not significantly changed, suggesting successful transfer to a more ideal substrate without significantly degrading graphene quality[12].

A four-point electronic measurement technique was used to measure the sheet resistance of the graphene film transferred onto a silicon substrate covered with thermally grown, 1.1-um-thick $SiO_2$. The conductance of a large-area graphene device ($1.1 \times 1.6 cm^2$) is plotted against gate voltage in Fig. 2f. The sheet resistance of layer-transferred graphene was found to be typically 400Ω (compared to 15Ω of our ITO films) at the room temperature. For single carrier transport, the carrier mobility is given by $\mu = \sigma/ne$, where $\sigma$ is the conductivity, $n$ is the carrier density, and $e$ is the elementary charge. Using the $SiO_2$/Si as a capacitive back gate, we vary the carrier density with an applied gate voltage $V_G$ and calculate the field effect mobility in our devices as $\mu_{FE}=(d/\varepsilon_0\varepsilon)(\partial\sigma/\partial V_G)$, where $d$ is the thickness of the $SiO_2$, $\varepsilon_0$ the permittivity of free space, and $\varepsilon$ is the dielectric constant (3.9 for thermally grown $SiO_2$). For the device in Fig. 2f, we measured $\mu_{FE} = 1,140$ $cm^2V^{-1}s^{-1}$. Slight non-linearity in Fig. 2f is likely due to a small but finite contact resistance in series with the graphene. We also found that both the resistivity and absorption of the layer transferred graphene decreased after vacuum annealing, likely due to the removal of adsorbed contaminants at elevated temperatures. We found similar effects when layer transferred graphene was illuminated with UV light, suggesting that the removal of the adsorbed contaminants was the most likely cause for the observed improvement in graphene quality.

*2.4 Solar cell fabrication and characterization*

For use in our hybrid photovoltaic devices, either graphene or ITO was deposited on glass and annealed at 140 °C for 10 minutes, followed by 15-minute exposure to UV light. PEDOT:PSS was spin coated and annealed at 200 °C for 30 minutes. The solution mixture of a P3HT:PCBM (1:1) polymer was then spin coated to form the active region of the photovoltaic device. The samples were subsequently baked in flowing Ar at 140 °C for 30 minutes in order to remove the solvent and crystallize the P3HT in the composite film. Finally, Al was deposited on top of the P3HT layer and used as a contact. Shadow masking was used to confine the device to a size of roughly 0.4×0.5cm$^2$.

Using a 532-nm continuous wave semiconductor laser as a light source, we measured the absorption dependence as a function of wavelength. As shown in Fig. 2g, the absorption of the graphene was roughly 1.5% in the visible range and decreased with increasing wavelength, similar to what was found previously[6]. After deposition, the P3HT:PCBM film featured an absorption of 42% at 540 nm (inset, Fig. 2g), similar to what was found previously[13]. Topographical characterization was also performed by AFM, shown in Fig. 2h with roughness about 1.3 nm.

## 3. Results and discussion

The current-voltage characterization of the hybrid devices fabricated from the photovoltaic composite was performed with a Keithley 4200 semiconductor parameter analyzer and a 532nm continuous wave semiconductor laser as the light source with a power intensity of 100mW/cm$^2$. All the measurements were carried out in air at room temperature. The *I-V* characteristics of the hybrid photovoltaic in dark and illuminated conditions are shown in Fig. 3a. Both graphene- and ITO-based devices showed similar values of open-circuit voltage ($V_{oc}$), ~0.62 V. The short-circuit density ($J_{sc}$) of graphene-based devices was found to be as high as 10.19 mA/cm$^2$, slightly better than the 9.74 mA/cm$^2$ measured in the ITO-based devices.

Similar behavior was found across multiple devices as shown in Table I. All devices have the same open-circuit voltage (0.62V), and a filling factor of around 25%. The effectiveness of a photovoltaic cell to convert incident photons of a fixed wavelength into photocurrent is measured by the incident monochromatic photon to current conversion efficiency (IPCE) given by[14,15]

$$ICPE = 1240 \frac{J_{SC}}{\lambda P_i}$$

where $J_{sc}$ is the short-circuit current density (A/cm$^2$), $P_i$ the incident photon power density (W/m$^2$), and $\lambda$ the excitation wavelength (nm). Our measurements suggest an IPCE of 2.38% and 1.98% for graphene- and ITO-based devices, respectively. For two ITO devices, the power conversion efficiency was found to be 3.49 and 3.86%. These numbers are lower than the best published data[16,17] for the material system used because the device structure and deposition conditions were not optimized. Nevertheless, the values of PCE for all graphene-based devices fabricated identically were found to be between 3.69 - 3.98%, competitive with and slightly higher than that of devices based on ITO.

Now we consider the factors limiting PCE in our layer transferred graphene photovoltaic devices. Different from conventional semiconductor devices, the dissociation of electron–hole pairs generated by the light absorption at the polymer:PCBM interface occurs because of the difference in the mobility of electrons and holes in the respective materials rather than the electrical field at a p-n junction. Transport of electrons through PCBM to the cathode and that of holes through the polymers to the anode are competing with the bimolecular recombination that results in the loss of photogenerated electron-hole pairs. This competition between charge transport and bimolecular recombination is the most important factor determining the PCE of the device. In addition, in devices presented in this study, the relatively low parallel shunt resistance of P3HT:PCBM films and contact resistance between PEDOT and layer transferred graphene may also play a role. It was shown previously that PCE of a conventional hybrid device could be improved by the inclusion of a thin LiF layer between the active layer and the Al electrode[18]. In the present work, the PCEs for both graphene and ITO devices measured under laser illumination using neutral density filters were found to decrease with increasing illumination intensity (Fig. 3b), most likely due to the resulting decrease in parallel shunt resistance[19]. The short circuit current density increases nonlinearly with power intensity (inset of Fig. 3b), indicating negligible bimolecular recombination in the device[20]. At high power intensity, the nonlinear rate appears to indicate a charge-density-dependent bimolecular recombination coefficient[20]. Both the charge density and recombination coefficient are influenced by charge trapping in the active layer. It is important to note that the substitution of graphene for ITO as a transparent electrode should not improve PCE in a photovoltaic device substantially. However, the comparative performances demonstrated by this work, along with graphene's characteristic advantages to ITO as outlined above, encourage the integration of graphene in future photovoltaics.

## 4. Conclusion

In conclusion, we demonstrated the fabrication of photovoltaic devices using layer-transferred graphene as a transparent electrode. Our layer-transferred graphene features a sheet resistance of about 400Ω, a field-effect carrier mobility of 1,140 $cm^2V^{-1}s^{-1}$ at room temperature and an optical absorption of 1.5%. The photon conversion efficiency and power conversion efficiency were found to be slightly higher in graphene-based devices than that of devices based on ITO. This, when considered along with graphene's remarkable flexibility, mechanical strength, and the possibility of low-cost production, suggests that CVD-grown graphene can be a competitive alternative for transparent electrodes in solar cells applications.

**Figure captions**

**Fig. 1.** (a) Schematic of the hybrid photovolatic device layout based on layer-transferred graphene; (b) Energy level diagram of a hybrid photovolatic device with graphene or $InTiO_3$ (ITO) as the transparent electrode.

**Fig. 2.** (a) Optical microscopy image of layer-transferred graphene on $SiO_2$/Si and (b) glass substrates; (c) Atomic force microscope (AFM) image of layer-transferred graphene on $SiO_2$/Si and (d) glass substrates; (e) Raman spectra of an as-grown graphene film on Cu foil and a layer transferred graphene film on a glass substrate, respectively; (f) Conductance versus back gate voltage of layer-transferred graphene on a silicon substrate covered with 1.107um of $SiO_2$. The red line highlights the linear regime along which the field effect mobility is calculated; (g) Absorption as a function of the wavelength for layer-transferred graphene (Inset: Absorption spectrum of P3HT:PCBM(1:1) film); (h) AFM image of a P3HT:PCBM film on glass.

**Fig. 3.** (a) Current-voltage character of the hybrid photovoltaic in dark and illuminated conditions using a 532 nm laser as a light source; (b) Power conversion efficiency as a function of the power density (Inset: Similar dependence of short-circuit density ($J_{sc}$)).

**Table 1.** Open-circuit voltage ($V_{oc}$), short-circuit current density ($J_{sc}$), filling factor, photon conversion efficiency, and power conversion efficiency of our photovoltaic devices with ITO and layer-transferred graphene as a transparent electrode.

Table 1:

| Sample name | $V_{OC}$ (V) | $J_{SC}$ (mA/cm$^2$) | FF (%) | ICPE (%) | Effic. (%) |
|---|---|---|---|---|---|
| ITO1 | 0.62 | 9.74 | 24.51 | 1.98 | 3.86 |
| ITO2 | 0.62 | 8.98 | 24.23 | 1.90 | 3.49 |
| LTG1 | 0.62 | 10.19 | 25.25 | 2.38 | 3.98 |
| LTG2 | 0.62 | 10.03 | 24.43 | 2.22 | 3.77 |
| LTG3 | 0.62 | 9.32 | 25.12 | 2.31 | 3.69 |

Fig. 1:

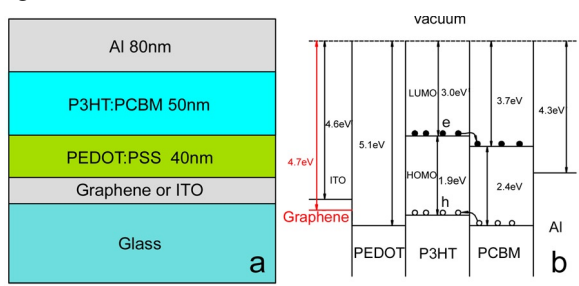

Fig. 2:

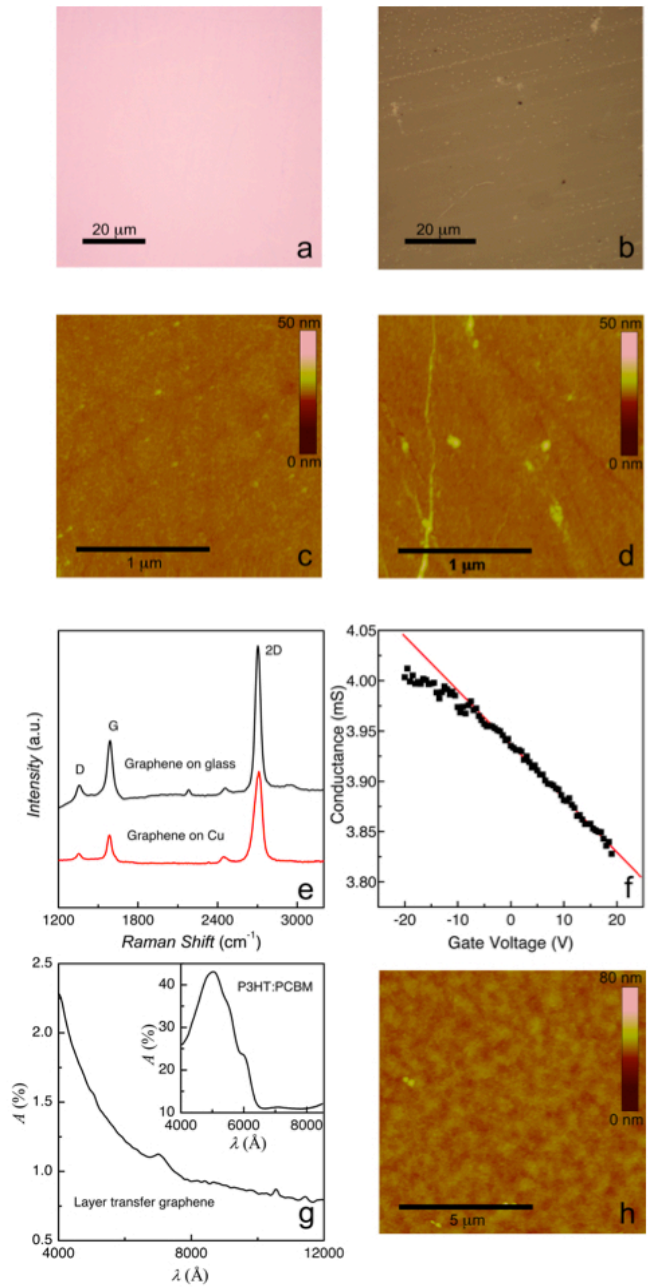

Fig. 3:

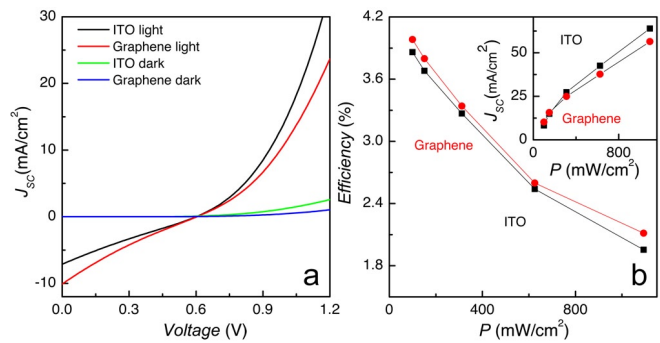